\documentclass[a4paper]{jpconf}
\usepackage{amssymb,amsmath,latexsym}
\usepackage{graphicx}

\usepackage{listings}
\usepackage{color}

\definecolor{dkgreen}{rgb}{0,0.6,0}
\definecolor{gray}{rgb}{0.5,0.5,0.5}
\definecolor{mauve}{rgb}{0.58,0,0.82}

\begin{document}
\title{NEUT/NuWro cross-section modelling at low three-momentum transfer}
\author{Patrick Stowell}
\address{University of Sheffield, Western Bank, Sheffield, UK}
\ead{p.stowell@sheffield.ac.uk}
\begin{abstract}
The MINERvA collaboration has reported a Charged-Current inclusive (CCinc) cross-section measurement on a CH target. This was performed by looking at both the muon and hadronic final state particles to create a double differential cross-section distribution that provides additional insight into the different regions of the phase-space where nuclear effects are present. We show early comparisons of the NEUT and NuWro generators in an attempt to estimate which parts of the models are in agreement with this data.

\end{abstract}
\section{Introduction}
Lepton scattering cross-sections have been found to differ significantly for light (H, D) and heavy nuclear (C, O) targets in both electron and neutrino beam experiments \cite{neutrinoreview}. These differences are caused by complex nuclear effects that vary with the target atomic number. Electron scattering experiments have found that a good seperation of these effects can be seen when measuring the cross-section in terms of the energy and three-momentum transfer $(q_0,q_3)$. Unfortunately, for neutrino scattering experiments it is difficult to directly measure  $q_0$ and $q_3$, since the incoming neutrino is not directly observed.
In a recent study \cite{minervadata} the MINERvA collaboration has tried to obtain a similar separation of nuclear effects in neutrino scattering through the measurement of hadronic final states. They measure the total energy deposited by different final state particles in a charged-current-inclusive sample and use this to form the kinematic quantity $E_{av}$,``Energy Available'', defined as
\begin{align}
E_{av} = \sum_{i=p,\pi^+,\pi^-} T^{Kinetic}_{i} + \sum_{j=e^{\pm},\gamma,\pi^{0}}  E^{Total}_{j}
\end{align}
This quantity has a direct correspondence with the energy transfer but depends on how nuclear re-scattering can modify the final state particle momenta. With this measure of $E_{av}$ and the muon kinematics an estimated $q_{3}$ is calculated so that events can be binned into a double differentical cross-section in $E_{av}-q_{3}$. In this distribution it has been found the the GENIE Monte-carlo (MC) generator \cite{geniegen} including the Valencia multi-nucleon (2p2h) interaction model \cite{nieves} is favoured over a model without multi-nucleon effects, but there are still tensions in the ``dip'' region between the quasi-elastic and resonance peaks.

\section{Generator Comparisons}
\begin{figure}
\centering
\includegraphics[width=0.75\textwidth]{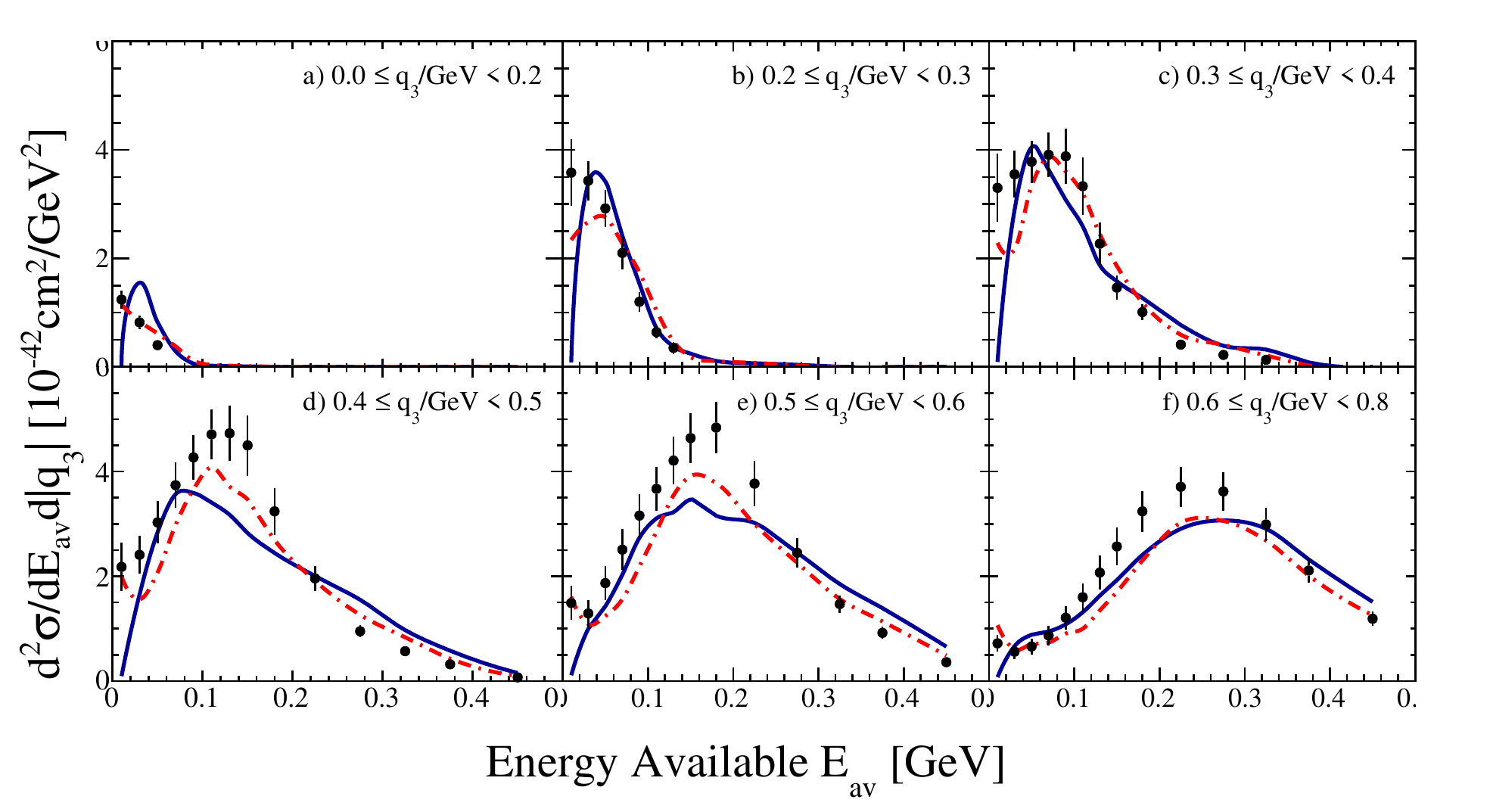}
\caption{\label{fig:ccinccomp} Model Predictions compared to MINERvA Low Recoil data in slices of three-momentum transfer. Shown are the published data distribution ($\fullcircle$), the NuWro generator prediction in red ($-\cdot-$), and the NEUT generator prediction in blue (\full).}
\end{figure}

To evaluate whether data/MC discrepancies are due to tensions in the underlying models it is important to compare new data to multiple MC generators and models. We choose two similar nominal models in the NEUT \cite{neutgen} and NuWro \cite{nuwrogen} event generators.
The full choice of model parameters can be seen in Table \ref{tabmodelchoice}.
\begin{table}
\caption{\label{tabmodelchoice}Details for the chosen nominal NEUT and NuWro models.}
\begin{center}
\begin{tabular}{lll}
\br
Generator & NEUT & NuWro \\
\mr
Nuclear Model & Relativistic Fermi Gas (RFG)\cite{smithmoniz} & Local Fermi Gas (LFG) \\
2p2h Leptonic Model & Nieves \cite{nieves} & Nieves \cite{nieves} \\
Resonant Model & Rein-Seghal (Full) \cite{reinseghal} & Rein-Seghal (Delta-only) \cite{reinseghal}  \\
FSI Model & Oset \cite{osetmodel} & Oset \cite{osetmodel}  \\
\br
\end{tabular}
\end{center}
\end{table}
For each model $2.5\times10^6$ events are generated on a CH target for all available interaction modes. The NUISANCE framework \cite{nuisance} is then used to select events that meet the true signal definition from the original MINERvA publication and a cross-section prediction is extracted. Comparisons of the total cross-section predictions to the data can be seen in Figure \ref{fig:ccinccomp}, where NuWro is found to have only a slightly better agreement than NEUT.

\section{Spectral Function Differences}
\begin{figure}[h]
\begin{minipage}{0.495\textwidth}
\centering
\includegraphics[width=0.95\textwidth]{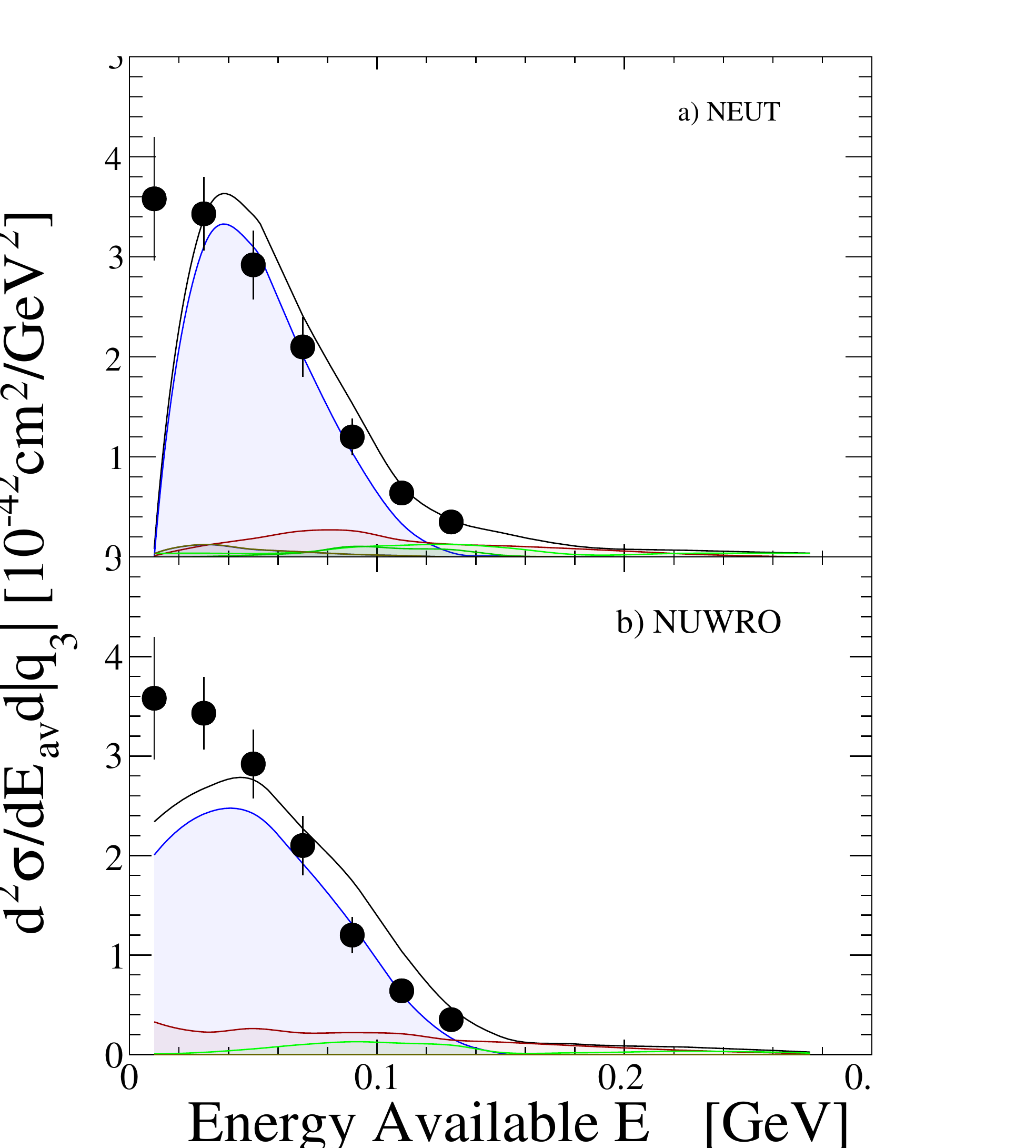}
\caption{\label{fig:loweav}CC inclusive cross-section prediction at low energy available where the quasi-elastic component (blue) dominates. Shown are the NEUT (a) and NUWRO (b) generators in the slice $0.2 \leq q_{3}/\mbox{GeV} < 0.3$. Shown in red is the 2p2h predicted contribution.}
\end{minipage}\hspace{2pc}
\begin{minipage}{0.495\textwidth}
\includegraphics[width=0.95\textwidth]{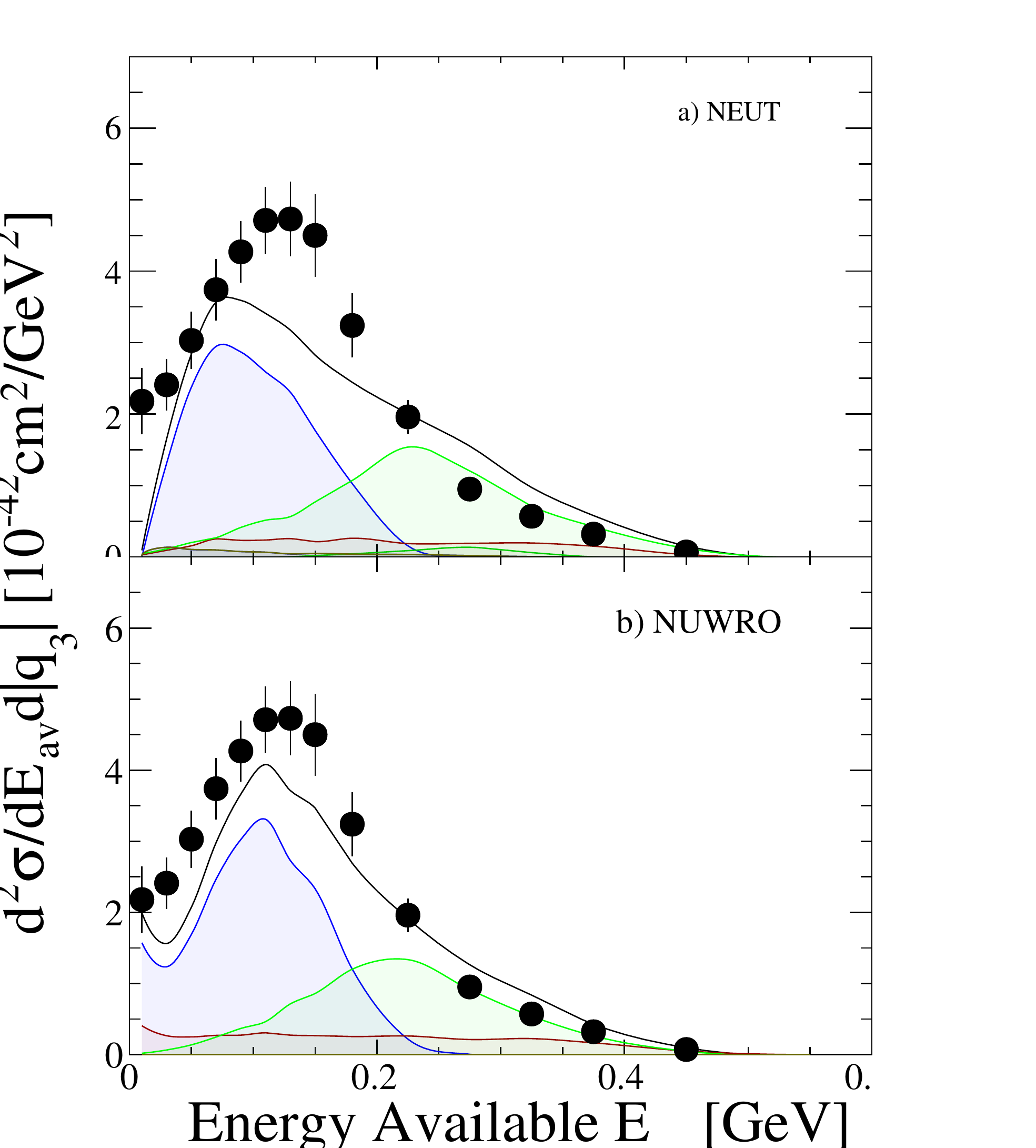}
\caption{\label{fig:higheav}Total cross-section prediction near the ``DIP Region'' between the quasi-elastic (blue) and resonant (green) contributions. Shown are the NEUT (a) and NUWRO (b) generators in the slice $0.4 \leq q_{3}/\mbox{GeV} < 0.5$. Shown in red is the 2p2h predicted contribution.}                                                                                                         
\end{minipage} 
\end{figure}

When comparing individual interaction channels we find the largest differences between the NuWro and NEUT generators are due to differences in their initial state models as expected. At low energy available and low three momentum transfer the events are dominated by purely proton/neutron final states. Here we see that the NEUT generator has a signficant deficit of events compared to NuWro, due to the exaggerated Pauli-blocking in the RFG model (Figure \ref{fig:loweav}). Unfortunately due to the difficulty in the direct tagging of low energy protons and the degeneracy in this region with other expected nuclear effects, multiple measurements will need to be made in future to place a good constraint on which spectral function best describes this region. At higher energies in the ``dip'' region (Figure \ref{fig:higheav}) there is a significant disagreement between the NEUT prediction and the data. This shape disagreement is found to be much smaller for the NuWro generator as the LFG model shifts the quasi-elastic peak to higher final state energies. Multiple cross-section channels compete here with the quasi-elastic and resonance peaks dominating, but since no direct measurement has been made on each of these channels exclusively in $E_{av}-q_3$, it is hard to determine whether the tensions come from the mis-modelling of a single interaction channel or multiple channels.

\section{Conclusions \& Future Work}
The tensions observed between NEUT and GENIE and the MINERvA $E_{av}-q_3$ data are seen to be smaller for NuWro when using a local Fermi gas model. The assumption that a single interaction channel is the cause of this tension builds on a larger assumption that the quasi-elastic and resonance channels are accurately modelled in the $E_{av}-q_3$ phase space. The best way to use these measurements to produce reliable model constraints is through a global fit to multiple cross-section measurements where all free parameters in the model are varied.

%\begin{figure}
%\centering
%\includegraphics[width=0.5\textwidth]{nu2016_proc_comp2_lowEav.pdf}
%\caption{\label{fig:loweav}CC inclusive cross-section prediction at low energy available where the quasi-elastic component (blue) dominates. Shown are the NEUT (a) and NUWRO (b) generators in the slice $0.2 \leq q_{3}/\mbox{GeV} < 0.3$. Shown in red is the 2p2h predicted contribution.}
%\end{figure}

%\begin{figure}
%\centering
%\includegraphics[width=0.5\textwidth]{nu2016_proc_comp3_dip.pdf}
%\includegraphics[width=0.5\textwidth]{nu2016_proc_comp2_lowEav.pdf}
%\caption{\label{fig:higheav}Total cross-section prediction near the ``DIP Region'' between the quasi-elastic (blue) and resonant (green) contributions. Shown are the NEUT (a) and NUWRO (b) generators in the slice $0.4 \leq q_{3}/\mbox{GeV} < 0.5$. Shown in red is the 2p2h predicted contribution.}                                                                                                         

%\caption{Total cross-section prediction near the ``DIP Region'' between the quasi-elastic (blue) and resonant (green) contributions. Shown are the NEUT (a) and NUWRO (b) generators in the slice $0.4 \leq q_{3}/\mbox{GeV} < 0.5$.}
%\end{figure}

\ack
The author wishes to thank the UK STFC funding council for supporting this work.

\end{document}